\documentclass[acmsmall]{acmart}

\usepackage{lineno}
\usepackage{todonotes}
\usepackage{comment}

\newcommand{\changed}{} 
\newcommand{\changedcamera}{} 
\AtBeginDocument{%
  \providecommand\BibTeX{{%
    \normalfont B\kern-0.5em{\scshape i\kern-0.25em b}\kern-0.8em\TeX}}}

\usepackage{todonotes}

\setcopyright{acmcopyright}
\acmJournal{PACMHCI}
\acmYear{2021} \acmVolume{5} \acmNumber{CSCW1} \acmArticle{116} \acmMonth{4} \acmPrice{15.00}\acmDOI{10.1145/3449190}

\begin{document}

%\title{Personalizing Social Comparison: An AI-Based Approach for Motivating Physical Activity}
%\title{Accuracy or Change: Lessons of Personalization Paradox from Personalizing Social Comparison for  Physical Activity}
%\title{Personalization Paradox in Adaptive Technologies for Behavior Change: \red{Lessons from Personalizing Social Comparison to Increase Physical Activity}}
%\title{Why Does the Personalization Paradox Escalate in Behavior Change Applications: \changedcamera{Personalizing Social Comparison to Increase Physical Activity}}
%\title{Personalization Paradox in Behavior Change Apps: Why Adaptation Based on User Models Is Circular Reasoning}
\title{Personalization Paradox in Behavior Change Apps: Lessons from a Social Comparison-Based Personalized App for Physical Activity}
%\title{Personalization Paradox: \red{Lessons from a Personalized Social m-Health App for Physical Activity/Social Fitness App}}
%\title{Personalization Paradox: Lessons from Personalizing Social Comparison for Physical Activity}
%\title{Personalization Paradox: Why Is it Difficult to Personalize Social Comparison for Motivating Physical Activity}

% \affiliation{%

\author{Jichen Zhu}\authornote{Currently at the IT University of Copenhagen, Denmark.}
\affiliation{%
  \institution{Drexel University}
  \city{Philadelphia, PA}
  \country{USA}}
\email{jichen.zhu@gmail.com}

\author{Diane H. Dallal}
\affiliation{%
  \institution{Drexel University}
  \city{Philadelphia, PA}
  \country{USA}}
\email{dd935@drexel.edu}

\author{Robert C. Gray}
\affiliation{%
  \institution{Drexel University}
  \city{Philadelphia, PA}
  \country{USA}}
\email{robert.c.gray@drexel.edu}

\author{Jennifer Villareale}
\affiliation{%
  \institution{Drexel University}
  \city{Philadelphia, PA}
  \country{USA}}
\email{jmv85@drexel.edu}

\author{Santiago Ontañón}\authornote{Currently at Google.}
\affiliation{%
  \institution{Drexel University}
  \city{Philadelphia, PA}
  \country{USA}}
\email{santi.ontanon@gmail.com}

\author{Evan M. Forman}
\affiliation{%
  \institution{Drexel University}
  \city{Philadelphia, PA}
  \country{USA}}
\email{evan.forman@drexel.edu}

\author{Danielle Arigo}
\affiliation{%
  \institution{Rowan University}
  \city{Philadelphia, PA}
  \country{USA}}
\email{arigo@rowan.edu}

\renewcommand{\shortauthors}{Zhu et al.}


\begin{abstract}
%planning document: https://docs.google.com/document/d/1x9EfSBh_WMojj8RzKMtoeK_eZqaRUjnhCIgkNPRrllo/edit 

Social comparison-based features are widely used in social computing apps. However, most existing apps are not grounded in social comparison theories and do not consider individual differences in social comparison preferences and reactions. This paper is among the first to automatically personalize social comparison targets. In the context of an m-health app for physical activity, we use artificial intelligence (AI) techniques of multi-armed bandits. Results from our user study (n=53) indicate that there is some evidence that motivation can be increased using the AI-based personalization of social comparison. The detected effects achieved small-to-moderate effect sizes, illustrating the real-world implications of the intervention for enhancing motivation and physical activity. \changed{In addition to design implications for social comparison features in social apps, this paper identified the {\em personalization paradox}, the conflict between user modeling and adaptation, as a key design challenge of personalized applications for behavior change. Additionally, we propose research directions to mitigate this Personalization Paradox.} %\red{rethink the m-health app emmphasis.}

\end{abstract}

\begin{CCSXML}
<ccs2012>
   <concept>
       <concept_id>10003120</concept_id>
       <concept_desc>Human-centered computing</concept_desc>
       <concept_significance>500</concept_significance>
       </concept>
   <concept>
       <concept_id>10003120.10003130.10003233</concept_id>
       <concept_desc>Human-centered computing~Collaborative and social computing systems and tools</concept_desc>
       <concept_significance>500</concept_significance>
       </concept>
 </ccs2012>
\end{CCSXML}

\ccsdesc[500]{Human-centered computing}
\ccsdesc[500]{Human-centered computing~Collaborative and social computing systems and tools}

\keywords{personalization, physical activity, social comparison, m-health}

\maketitle

% ACM Classfication

% <concept_id>10003120.10003121</concept_id>
% <concept_desc>Human-centered computing~Human computer interaction (HCI)</concept_desc>
% <concept_id>10003120.10003121.10003125.10011752</concept_id>
% <concept_desc>Human-centered computing~Haptic devices</concept_desc>
% <concept_significance>300</concept_significance>
% <concept_id>10003120.10003121.10003122.10003334</concept_id>
% <concept_desc>Human-centered computing~User studies</concept_desc>
% <concept_significance>100</concept_significance>

% \ccsdesc[500]{Human-centered computing~Human computer interaction (HCI)}
% \ccsdesc[100]{Human-centered computing~User studies}

%--------------------------------------------
% INTRODUCTION
%--------------------------------------------

\section{Introduction}
%\todo[inline]{Jichen}
%What is the broader issue that your project will address [CONTEXT/MOTIVATION]

%
\changedcamera{
Personalization technology has been adopted in a wide range of digital applications such as health, training and education, e-commerce, and entertainment~\cite{zhu2020Personalization,snodgrass2019like}. By using artificial intelligence (AI) to tailor themselves to individual users' needs and preferences, personalized systems have shown to improve learnability~\cite{furqan2017learnability}, usability~\cite{hook1998evaluating}, and user enjoyment~\cite{shaker2010towards}. }

An active area of personalization is technologies for behavior change, especially for health or learning. In digital health and fitness applications, researchers have used personalization to establish and tailor individualized physical activity goals for improved health benefits~\cite{rabbi2015mybehavior, yom2017encouraging,sockolow2017risk,zhou2018evaluating}. By a recent account, the global market of fitness apps has reached \$3.3 Billion in 2019 and is expecting rapid growth in the next few years.\footnote{retrieved from https://www.reportlinker.com/p05881751/Fitness-App-Market-Research-Report-by-Function-by-Type-Global-Forecast-to.html on Oct 1, 2020} The potential societal benefits for personalization on these platforms is significant.

% Approximately two-thirds of the U.S. adult population is overweight or obese \cite{ogden2012prevalence}. %More alarmingly, obesity among young people is rapidly growing. 
% A main contributing factor to this problem is sedentary behavior. 
% %According to a study, only 8\% of adolescents and young adults get the recommended 60 minutes of daily physical activity (PA) \cite{troiano2008physical}. 
% %
% An increasingly popular approach for increasing physical activity (PA) is m-health fitness apps. By a recent account, the global market of fitness app has reached \$3.3 Billion in 2019 and is expecting rapid growth in the next few years\footnote{retrieved from https://www.reportlinker.com/p05881751/Fitness-App-Market-Research-Report-by-Function-by-Type-Global-Forecast-to.html on Oct 1, 2020}. 

This paper focuses on applying personalization to the social features commonly used in fitness apps. Recently, many commercially available fitness apps for physical activity (PA), such as {\em Fitbit}, {\em Endomondo}, and {\em Strava}, use social features to increase user engagement and motivate PA. They allow users to make friends with each other, post their PA experiences (e.g., performance data and exercise route), and participate in communities with common fitness goals. %Some of the most common features are ranked leader boards and competitive challenges~\cite{conroy2014behavior,arigo2020social}. %For instance, {\em Fitbit} {\em StepUp} and {\em Endomondo} have a leader board of steps and allow friends to challenge one another. {\em Strava} and {\em Endomondo} enables users to upload their exercise information. 
Some of the most widely used social features are ranked leader boards and competitive challenges~\cite{conroy2014behavior,arigo2020social}.
The key psychological process that underlies these features is {\em social comparison}, which is described in social psychology as the fundamental psychological process by which individuals evaluate themselves or their behavior relative to others~\cite{festinger1954theory}.

Despite the wide usage of social comparison-based features, most existing social fitness apps are not grounded in evidence from recent psychology theories of social comparison~\cite{arigo2020social}. In particular, current research shows that individuals differ in the direction in which they prefer to compare themselves (i.e., comparing {\em upward} with those who are ``better off'' versus comparing {\em downward} with those who are ``worse off''). Furthermore, when comparing to targets of their preferred direction, individuals may react positively or negatively, both in terms of the desired behavior (e.g., PA) and their motivation for it.  To date, most (if not all) digital PA interventions do not address these known individual differences~\cite{arigo2020social}. With very few exceptions~\cite{mollee2016effectiveness, klein2017active2gether}, little research has been done in personalizing the social comparison targets provided to individual users. We believe that grounding the design of social fitness apps in current evidence from psychology and providing personalized social comparison opportunities can increase the apps' effectiveness to motivate a wide range of users to be more active. 

This paper presents a novel approach to personalizing social comparison automatically. We designed an Artificial Intelligence (AI)-based web app that 1) models individual users' real-time preference of social comparison direction and their reaction to the comparison and 2) adapts the pool of comparison targets accordingly. For individual users' reactions to social comparison, we include their behavioral reaction (in terms of their daily step counts) and motivational reaction (their reported motivation to exercise). To evaluate our approach, we conducted a 21-day user study (n=53) to investigate our personalization mechanism and how exposure to personalized social comparison targets may affect users' PA and motivation to exercise. %Our study led to promising results and shown that AI-based personalization of social comparison has the potential to increase users' PA and their motivation for exercise. More specifically, 
The study results indicate that our AI-based personalization approach was able to automatically model and manipulate social comparison in the pursuit of PA promotion. The detected effects achieved small-to-moderate effect sizes, illustrating the real-world implications of the intervention for enhancing motivation and PA. 

\changed{Reflecting on this work, we identified a fundamental conflict in the design of personalization technology: the {\em Personalization Paradox},\footnote{\changedcamera{The term ``personalization paradox'' occasionally serves as a shorthand for the ``privacy-personalization paradox''~\cite{aguirre2015unraveling} (personalization creates users' sense of vulnerability and lower adoption rates), which is different from the use in this paper.}} which occurs when personalization adapts the digital environment based on its model of a user. In doing so, it also changes the user, and thus the original user model is no longer accurate. This paradoxical relationship between user modeling and adaptation is particularly acute in technologies for behavior change. 
%
%personalization technology and its applications for behavioral change such as personalized health intervention, personalized learning (both are active research areas with many commercial applications).
%
Based on our study, we offer initial suggestions to mitigate this personalization paradox.}

\changedcamera{For precision and consistency, we clarify our terminology for the rest of the paper. Personalization algorithms usually consist of two steps: {\em user modeling} and {\em adaptation} of the digital environment. In the rest of this paper, we will use the term ``personalization'' to refer to the whole process, and ``adaptation'' to refer exclusively to the second step.}

The main contribution of this paper is threefold: 
\begin{itemize}
    \item First, to the best of our knowledge, this is the first attempt to {\bf automatically personalize for social comparison}, a widely adopted feature for social m-health apps, and empirically assess its effect on users' motivation and PA. Compared to related work built on users' self-report~\cite{mollee2016effectiveness,klein2017active2gether}, we used AI to automatically infer users' preference and reaction. Our approach to model user's social comparison using the AI technique of multi-armed bandits (MAB) can be applied to other personalization social computing applications. 
    \item \changed{Second, our study generated new evidence that can advance the {\bf scientific understanding of human social comparison processes}. While the established psychology literature treats social comparison as a stable individual difference, our study gathered more fine-grained data on users' social comparison behavior over 21 days. We discuss how our data may provide new evidence to further psychology research on social comparison as a dynamic process, which may improve how we design social computing, as social comparison is a ubiquitous psychological process we all engage in.} 
    \item \changed{Third, perhaps most important, reflecting on our process and results, we identify a less-understood challenge in developing personalization systems for behavioral change --- the {\bf personalization paradox}. We extend the original concept by identifying two main causes: the self-reinforcing loop problem and the moving target problem.  %In addition to self-reinforcing tendency of personalization technology and its negative social implications ~\cite{o2016weapons,noble2018algorithms,pariser2011filter}, we identify the personalization paradox as another key challenge to design effective and ethical personalization systems. 
    Our reflection and initial suggestions on the topic can help designers of personalization systems mitigate the personalization paradox.} 
\end{itemize}

The rest of the paper is organized as follows. We first review related literature on personalization digital interventions for PA and introduce the theory of social comparison. Next, we describe our personalization m-health app for PA. We then describe the methodology of our user study and discuss our results. Finally, we offer our interpretations of our results and design guidelines. 

\section{Related Work}
This section presents related work in personalization digital intervention for PA, current theory on social comparison, and social comparison-based digital interventions for PA. 

\subsection{Personalization of Digital Interventions for PA}

Among efforts to promote PA, there is a broad base of existing literature on personalized interventions, especially to adapt PA goals for improving goal adherence~\cite{parkka2010personalization,bleser2015personalized,bull1999effects, rabbi2015automated,arigo2020social}. %For example, previous interventions have personalized physical activity goals and motivated goal adherence based on a wide range of factors, including one's common activity types (e.g., running vs. cycling)~\cite{parkka2010personalization}; frequency, intensity, and duration of activity~\cite{bleser2015personalized}; visualized and message-based feedback on tracked data, along with progress towards behavioral goals~\cite{bull1999effects, rabbi2015automated}; and activation of social comparison to other individuals performing better and/or worse than a given user \cite{arigo2020social}.
Among existing work on PA personalization, an emergent approach is to use artificial intelligence (AI) techniques to automatically personalize digital interventions~\cite{paredes2014poptherapy,yom2017encouraging,forman2019can}. A benefit of AI-based interventions is the ability to provide real-time tailoring of feedback and adaptation of goals based on continuous data monitoring on an individual basis. 
%Although physical activity personalization has been investigated in depth, prior research on personalized mobile interventions which use artificial intelligence (AI) systems to promote physical activity has been limited thus far. AI-based personalization is a valuable tool for facilitating physical activity, given its unique ability to provide real-time tailoring of feedback and adaptation of goals based on continuous data monitoring on an individual basis. For these reasons, AI-based personalized approaches are also critical tools for health behavior change interventions beyond physical activity promotion, and have begun to be evaluated in such contexts. 

Extant research on health behavior change interventions has found that personalization AI systems are more acceptable to users than generic automated systems due to the relevance of health behavior recommendations delivered~\cite{rabbi2015mybehavior}. %In addition, personalization AI systems have been shown to reduce momentary stress \cite{paredes2014poptherapy}, improve glucose levels in diabetic adults \cite{yom2017encouraging}, and produce equivalent weight loss outcomes to non-optimized treatment while reducing intervention costs for overweight and obese adults \cite{forman2019can}. 
Most importantly, the few personalization AI systems designed to enhance PA specifically have been shown to increase PA, reduce calorie intake over time, and help individuals develop more challenging but attainable step goals \cite{rabbi2015mybehavior, yom2017encouraging, zhou2018evaluating}. %These AI-based systems for physical activity promotion personalization their interventions based on automatic and manual user inputs including: number of minutes of activity in the last day, cumulative number of minutes of activity in the week, proportion of activity goal achieved, proportion versus expected goal achieved at a given point in the week, age, and gender. In addition, one simulation study used each user's ``context'' (i.e., location, number of app screens viewed in the previous day, number of previous push notifications, and step count variation) to personalize the subsequent activity suggestions~\cite{liao2020personalized}. Results from the simulation study indicated that when context-tailored activity suggestions were delivered, 29 of 37 simulated participants experienced greater reward (i.e., higher reward scores) and behavioral benefit (i.e., greater step count 30 minutes after feedback was delivered). Thus, although the research thus far has been limited, there is preliminary evidence that personalization AI systems to promote physical activity are acceptable and engaging to users, and facilitate improvements in physical activity over time when personalizing based on demographic characteristics, activity history, and degree of goal achievement. 

%{\color{red}[Good summary of AI-based intervention in Psyc literature. We need to find more on the HCI/AI side. Jen, Gray, can you help?]}
With the wide adoption of wearable sensor technology and smartphones, much work on personalization in health applications in the HCI literature has focused on designing what information to provide the user at exactly the best time~\cite{akker2014tailoring}. For example, previous systems have personalized the content and timing of activity recommendations based on a wide range of factors, including demographic information~\cite{buttussi2008mopet}, behavioral patterns~\cite{lin2012bewell}, time and location~\cite{stahl2008mobile, buttussi2008mopet}, personality~\cite{arteaga2010mobile}, social comparison behaviors~\cite{klein2017active2gether, king2013harnessing}, and the user’s social community~\cite{lin2012bewell, king2013harnessing}.

\changed{Critically, relatively little work has been done to evaluate the effectiveness of {\bf AI systems designed to personalize psychological processes}, which are the underlying mechanisms of change that may lead to increased PA. Rather, the literature on AI systems to enhance PA has largely focused on establishing and tailoring PA goals.} One promising psychological process to target in the context of PA tailoring is social comparison (as described below). Despite the wide adoption of social comparison-based features and the central role they play in digital apps for PA, only a handful of existing work explores how to personalize the social comparison environment~\cite{klein2017active2gether, king2013harnessing}. This paper thus extends the literature on personalization in digital interventions for PA by further developing how to design and develop applications to personalize social comparison as a mechanism to enhance PA. \changed{In this paper, we do not directly attempt to tailor PA goals, because establishing an AI system's ability to engage a psychological process that precedes PA is a critical first step that is missing in the current literature.} %\todo[inline]{Dani: can you help me come up with a reason here? -- Diane added some revisions in red.}

\subsection{Social comparison}
Social comparison describes the process by which individuals evaluate themselves or their behavior relative to others~\cite{festinger1954theory}. %This is thought to be innate and protective, as it can allow individuals to assimilate to social circumstances and avoid standing out in negative ways.
Comparisons can happen quickly and automatically when an individual is exposed to information about other people~\cite{gilbert1995comparisons}, or they can result from more conscious decision-making processes (e.g., selecting a comparison target from among many choices, as in a group setting~\cite{arigo2014social}).
Social comparison processes in intervention groups have been shown to motivate health behavior change in the contexts of weight control and PA promotion~\cite{olander2013most, leahey2010effect}. In both commercially available m-health products and research prototypes,  social comparison and social support are considered to be the major mechanisms to affect PA~\cite{lin2012bewell,arigo2020social,lewis2015using, conroy2014behavior}. Social comparison-based features such as leader boards and competitive challenges have been used in many m-health systems~\cite{lin2012bewell,arigo2020social}. However, as explained below, most existing apps do not account for individual differences and provide all users with the same kind of social comparison. As explained below, this design assumption is not completely aligned with current theories of social comparison. 

 %, when individuals hear about group members' achievements or lapses and see changes in their physical appearance (e.g., body size or muscularity). 
%With respect to digital interventions, social comparison is often prompted in commercially available physical activity smartphone apps by features such as ranked leader boards and competitive challenges~\cite{conroy2014behavior,arigo2020social}.

Research has shown that individuals prefer different {\em social comparison targets}~\cite{bennenbroek2002social}. Comparisons to others who seem ``better off'' than the comparer in a given domain are {\em upward comparisons}, and comparisons to others who seem ``worse off'' are {\em downward comparisons}~\cite{buunk1995comparison}. Despite their preferences, individuals' responses to different social comparison targets vary~\cite{van1999big, van2000social}. Upward and downward comparisons %on any scale 
can motivate behaviors such as PA if the comparer sees an upward target's status as achievable for themselves or a downward target's status as one to be avoided~\cite{buunk1997social}. The opposite is also possible; an upward target's status may be seen as unreachable, serving only to highlight the comparer's inferiority, and a downward target's status may be seen as an indication that one's own situation is already satisfactory, serving to negate any need for behavior change efforts~\cite{arigo2015addressing, merchant2017face, wills1981downward}. Furthermore, an individual's preferences and responses may vary over short periods~\cite{arigo2020social}. For this reason, in part due to this range of possible outcomes, the current approach used by most (if not all) commercially available digital PA interventions does not address these known individual differences and hence is inadequate~\cite{arigo2020social}.

A small number of research prototypes have attempted to personalize social comparison. Mollee and Klein~\cite{mollee2016effectiveness} used a self-reported social comparison preference to adapt users' comparison targets, where they found that their approach was more effective than generic exposure. Klein et al.'s~\cite{klein2017active2gether} {\em Active2Gether} system, a personalized m-health intervention, had built-in social comparison at both individual and group levels. Similar to Mollee and Klein~\cite{mollee2016effectiveness}, {\em Active2Gether} asked a user's preferred comparison direction in an intake questionnaire and showed the user six comparison targets in their preferred direction. 

Our project extends existing literature by automatically detecting users' social comparison preferences based on user behavior without only relying on self-report, which may be incomplete with respect to how well preference relates to desired behavior change. %Response bias in self-assessment is widely discussed in behavioral science. 
Our approach can be complementary to the self-assessment-based work. Additionally, instead of only considering users' preference of comparison direction as in previous studies~\cite{mollee2016effectiveness,klein2017active2gether}, our approach also takes into account users' social comparison reactions by monitoring changes in their PA and their motivation to exercise.

%shown that simple attempts to tailor physical activity comparison targets to individuals' comparison preferences is more effective than generic exposure~\cite{mollee2016effectiveness}. Yet, work in related domains also demonstrates that merely allowing individuals to choose their targets can have negative effects~\cite{buunk1997social, schokker2010impact} (consistent with the explanation above), and that selected targets actually work against the comparer's intentions (e.g., leading to decreases in motivation or confidence in behavior change, rather than the intended increases~\cite{arigo2018social}). Furthermore, single indications of preference do not capture the spectrum of comparison processes (e.g., how many or what combination of targets might be desired~\cite{van1998social}) or address potential short-term variation in preference (e.g., due to changes in mood or the comparer's own status). To fully optimize the potential of tailoring social comparison opportunities to promote physical activity, it will be necessary to address both preference and short-term consequences of comparison, such as motivation and actual activity engagement.

\section{Designing A Social Comparison-Based Personalization Platform} 

To investigate how to personalize social comparison for PA, we designed a web-based platform in which users can compare themselves, including their daily steps, with other users' PA-related profiles. The user's steps are captured by {\em Fitbit} and synced automatically with our platform. %Before using it, the user is asked to sync her Fitbit with the website so that it receives the user's steps info. 
We use the AI technique of multi-armed bandits to model individual users' social comparison preferences and adapt the comparison targets shown to them. 
\changed{We selected steps as the indicator for PA because a person's daily step count is a widely recognizable and intuitive variable, and it is therefore likely to appeal to a large subset of the general population. We did not include other PA-related measures such as the intensity of PA, because the most recent evidence suggests any type of PA for any length of time has benefits~\cite{jakicic2019association}, which is now reflected in the national guidelines for PA~\cite[pp.110]{piercy2018physical}. Future studies can examine effects on parameters such as sedentary time, moderate-to-vigorous intensity activity, and other biometric measures.}

\begin{figure}[t!]
\includegraphics[width=0.85\columnwidth]{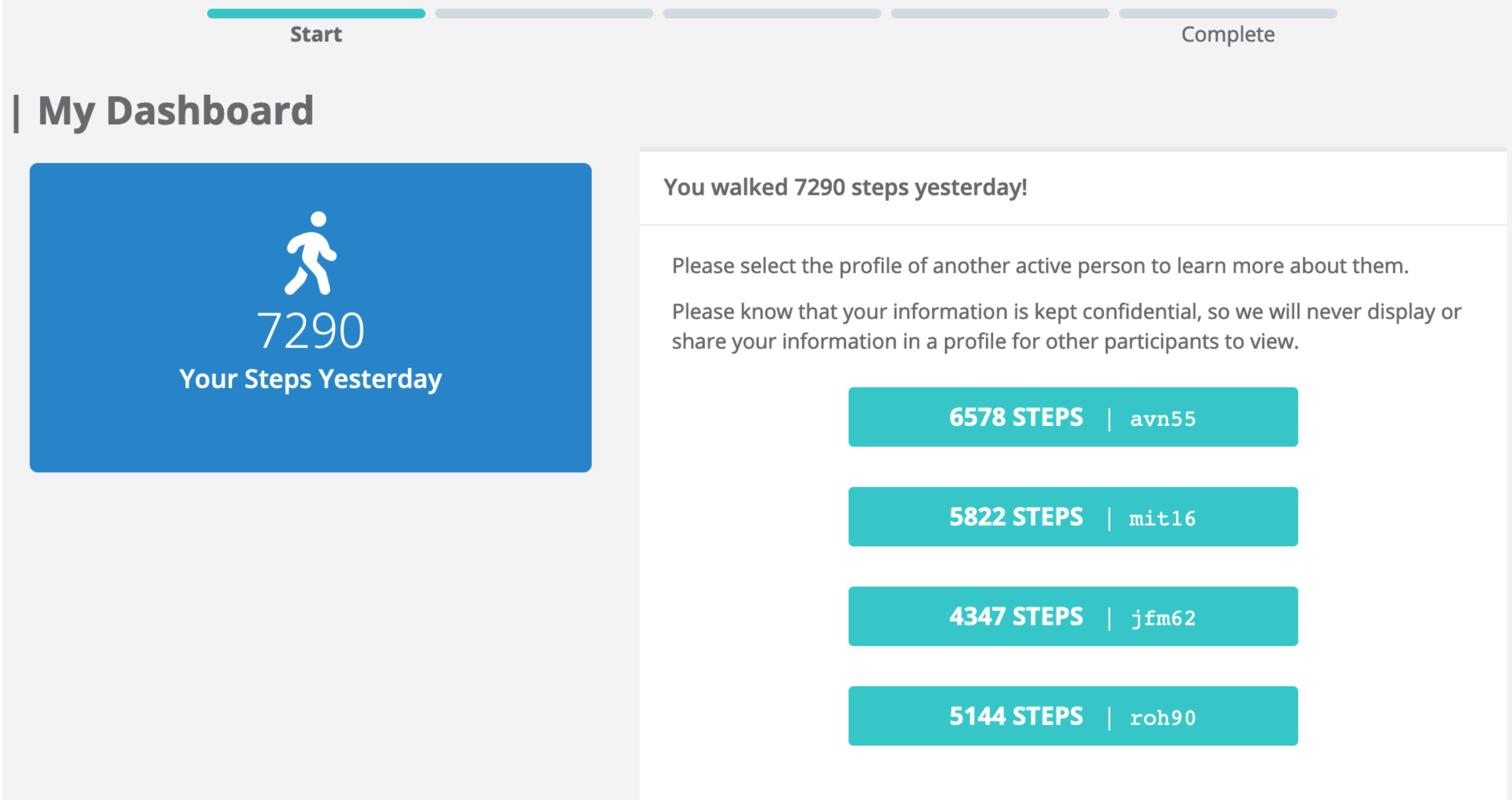}
\centering
\caption{Screenshot of the User Profile Selection Page.}
\label{fig:choicepoint_1}
\end{figure}

%MOve to methodology
%2.3.1. Pre-survey
%When a user logs into our website, she is first asked to answer to the question ``Overall, I would rate my current motivation to exercise as.... [very low / somewhat low / neutral / somewhat high / very high].

\subsection{Interaction with User Profiles} \label{section_profileSelection}
In this work, we used artificially generated user profiles to easily manipulate social comparison conditions. However, the principles of our design can be generalized to real users when used in fitness apps with a social network such as in many commercially available apps and research prototypes ~\cite{king2013harnessing,klein2017active2gether}. A user's interaction with the web app consists of the following three steps. 

{\bf {\em User Profile Selection.}} Each day when a user logs in, the app displays the user's daily step from the previous day (automatically retrieved from the user's {\em Fitbit} account) and four new user profiles consisting of those profiles' respective non-descriptive user names (e.g., ``azb30'') and total steps. We purposefully designed this User Profile Selection page with minimal information so that the user could focus on a single dimension for social comparison: the previous day's total steps. The user will see different user profiles every day, all of which are generated by artificial intelligence (AI) (details in Section \ref{section_AI}). The user is informed that she can preview all profiles but only review one {\em full} user profile each day.  \changed{This design follows established selection methods in social comparison literature~\cite{wood1996social,arigo2014social,gerber2018social}. It also intends to provide information about participants’ decision-making about the choice (e.g., how many profiles were selected before the final, full profile was chosen) while minimizing the time burden of daily participation.}

%We chose to use the previous day's step data, instead of that of the current day, so that users who log in early in the day without many changes to accumulate steps will not be discouraged. 
%

%which changes every day. In each profile, the only information provided at this stage is the total steps of the day before. We decided to use  non-descriptive user name (e.g., ``azb30''). This page is designed purposefully with minimal information so that the user can focus on the information for social comparison - the previous day's steps. The user is told that she can only review one user profile each day. 

\begin{figure}[t!]
\includegraphics[width=0.85\columnwidth]{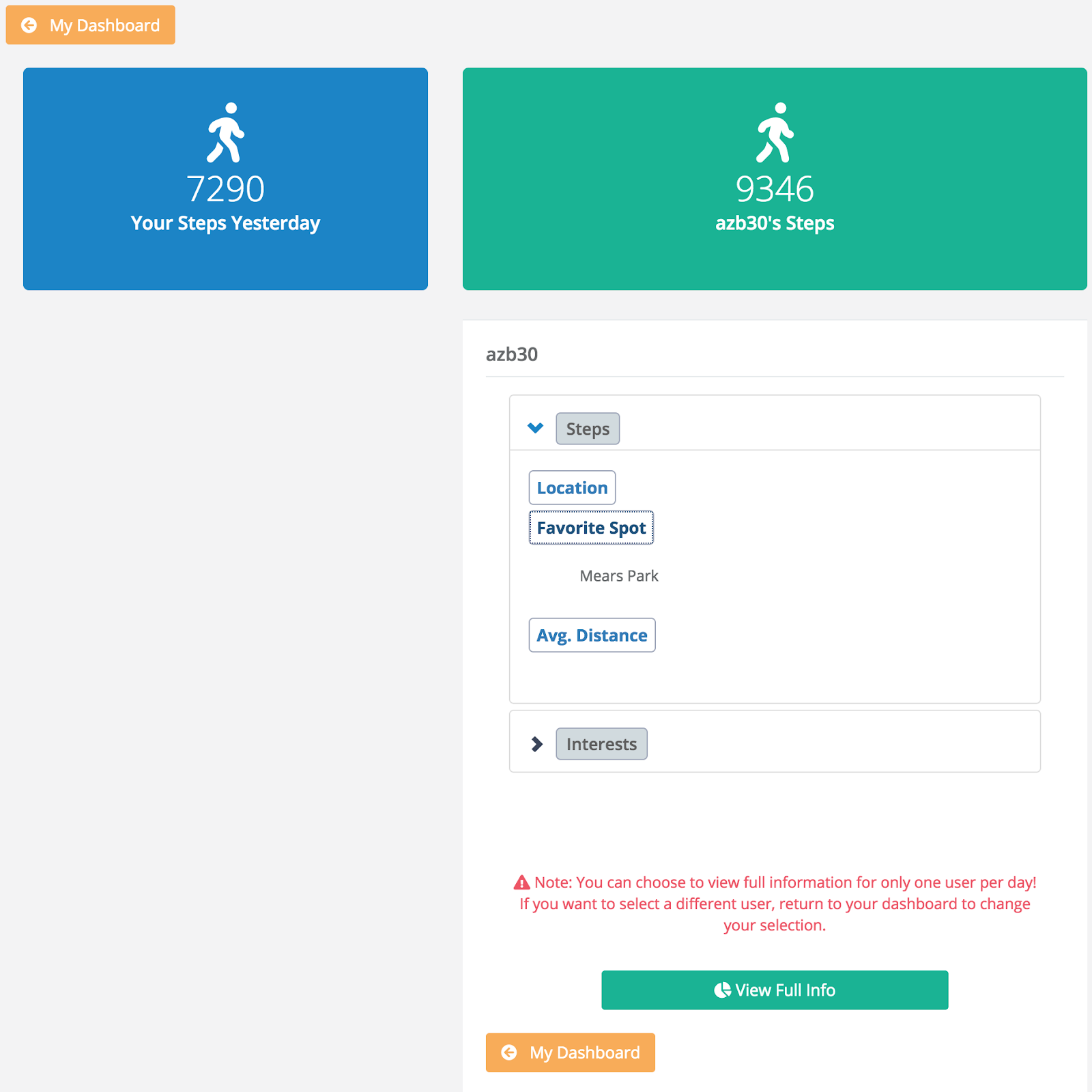}
\centering
\caption{Screenshot of the Overview Page, showing the user's steps (left) and the overview of a user profile (right). The dropdown menu of ``Steps'' is expanded, and the field ``Favorite Spot'' is further expanded.}
\label{fig:choicepoint_2}
\end{figure}

{\bf {\em Selection Overview.}} After selecting a user profile in the previous step, the user will see an overview of the selection (Fig. \ref{fig:choicepoint_2}). This page displays two key categories of user information: steps and interests. Notice that the user must click on the drop-down menu of each category to see more information (See the expanded ``Steps'' menu in Fig. \ref{fig:choicepoint_2}). 
We deliberately require multiple clicks for the user to unlock information to track which information, if any, the user is paying attention to. The user is reminded that she can only view one full user profile a day. The user may return to the previous Selection Overview page for the overview of different profiles. 
%This design allows the app to track what the user has shown interest. %For instance, expanding ``Steps'' will reveal three subcategories. Clicking any subcategory, the user can see the specific information. 

%Choice Point 2 (Overview): After choosing a user’s profile, a user is given the overview information of the selected user profile. Compared to Choice Point 1, the user is given more  information about two categories: ``Steps'' and ``Interests.'' 

{\bf {\em Full Profile.}} After the user confirms her selection, she is given more detailed information about the selected user profile. The Full Profile reveals further information such as demographic data (e.g., age, sex, profession), health information (height, weight), exercise preferences (average time spent at the gym, preferred physical activities), and other personal information (hobbies).

\changed{
A key design decision was to remove design elements that may influence users' social comparison processes. Notice that we did not rank users based on their daily steps (Fig. \ref{fig:choicepoint_1}), which is a design feature used in related work \cite{mollee2016effectiveness,klein2017active2gether}. We also refrained from setting step goals. This is because constructs such as competition, goals, and winning may create incentives for upward comparison and thereby skew users' innate social comparison preference. We will discuss the implications of this decision below.} 

%
% Move to Methodology
% 2.3.3. Post-survey
% After interacting with the user profiles, the user will complete a daily post-survey with two questions. First, she is asked ``Overall, I would rate my current motivation to exercise as.... [very low / somewhat low / neutral / somewhat high / very high]. 
% Finally, she is asked ``Was yesterday a typical day for you, in terms of your physical activity (number of steps)? [More than a typical day     / A typical day     / Less than a typical day]     

\subsection{Personalization Mechanism} \label{section_AI}
%\todo[inline]{Gray}

Our app personalizes users' social comparison targets by automatically modifying the user profiles presented to them on a daily basis. In particular, our AI 1) models each user's social comparison preference (i.e., upward or downward) and 2) adapts the steps of the user profiles shown to the user accordingly. Below we describe the personalization mechanism we designed at a high level. We build on the technical foundations described in recent work~\cite{gray2020player,gray2020cog}.  

We used the AI technique of multi-armed bandits (MABs)~\cite{Auer2002, kuleshov2000, lattimore2018bandit}
%~\cite{thompson1933likelihood, robbins1952some} 
to model users' social comparison preference and to personalize the user profile options. Multi-armed bandits are useful to model situations in which a system needs to repeatedly choose one among a series of options trying to find the one that gives it the maximum reward. Each time the system selects one option, a reward is observed, and over time, the system learns which of the options achieves the highest expected reward. Various AI techniques have been used for user modeling, such as sequential machine learning~\cite{valls2015exploring} and supervised learning~\cite{kantharaju2019tracing}. %\red{We chose MABs, instead of other AI techniques, because ....} %\todo[inline]{Santi: please add a couple of sentences, preferably with citations to back up your claim. Reviewer asked us to ``also mentioned other AI techniques and provided more reasoning why choose Multi-armed bandits over other those techniques.''}
\changed{
In our case, two factors make MABs particularly suitable. First, we cannot assume that participants in real-world conditions will have consistent behavior; the same person may not choose the same comparison target each time, even given identical profiles. Second, to discover the best selection of social comparison targets (user profiles), showing users other potentially worse options is necessary. This is commonly known as the exploration/exploitation dilemma. Solving this exploration/exploitation dilemma under uncertainty is precisely the problem that MAB strategies were designed to solve~\cite{Auer2002}. MABs lie at the core of other AI techniques such as Reinforcement Learning and Monte Carlo Tree Search. Given the amount of data we anticipated to collect from our study, we chose to use the core implementation of MAB strategies, which can be seen as Reinforcement Learning with a single state.}

When a user logs in to our app, the AI chooses one among three options---called an MAB ``{\em arm}''---to model the user's social comparison preference: {\em Downward}, {\em Mixed}, and {\em Upward}. Depending on which arm the AI selects, four different user profiles will be shown to the user. \changedcamera{In the literature, AI techniques have been developed to modify or generate text-based digital content~\cite{zhu2010story,zhu2010towards}. In our system, the user models (e.g., a selected MAB arm) modifies the social comparison targets a user will be exposed to.} According to Buunk's theory~\cite{buunk1995comparison}, social comparison targets may be perceived as only slightly better/worse off than the self, moderately better/worse off, or extremely better/worse off, with variations in between~\cite{buunk1995comparison}. We thus designed the generation of step counts for artificial user profiles relative to the user's performance accordingly (Table \ref{tab:mab-arms}). For example, if the {\em Downward} arm is selected by the AI, the four profiles shown will all have lower daily step counts than the user, respectively showing 40\%, 30\%, 20\%, and 10\% fewer, with a random factor (within $\pm$2\%) applied for obfuscation, and displayed in random order. 

The reward function used by the MAB algorithm to adjust itself is an equal combination of the changes in user's self-reported motivation for exercise before and after using the app (see Section \ref{sect:methods}) as well as her total steps during that day. Based on the previous days' rewards, the MAB algorithm determines which arm will most likely maximize this reward for the current day. In other words, in our work the AI learns which types of social comparison targets, instantiated through user profiles, are more likely to motivate a specific player to engage in PA. It is important to note that our AI does not encode any domain knowledge. For instance, consistent with existing literature on social comparison-based interventions~\cite{mollee2016effectiveness,klein2017active2gether,o2009upward,mahler2010effects,leahey2007frequency}, we assume a person prefers either the ``upward'' or ``downward'' direction. However, our AI does not know, for example, that the Mixed arm is an intermediate option for when this preference is uncertain, nor does it bias towards any one of the three arms. 

%while in psychological research on social comparison, it is conventional to consider a person either an ``upward'' or ``downward'' comparer. 

%As the feedback to its choice of the three arms (i.e., social comparison preferences), the AI records the change in the user's self-reported motivation for exercise before and after using the app (see Section \ref{sect:methods}) as well as her total steps during that day. Based on those two data points, a numeric reward is calculated to determine which one of the three arms to choose the following day to maximize this reward. In other words, by using the MAB, the AI learns which types of social comparison targets, instantiated through user profiles, motivate specific player the most to do physical exercise.  %Based on this idea, we designed an MAB-based AI approach~\cite{gray2020player,gray2020cog} for our app where the MAB chooses which types of user profiles to show each user every day in order to maximize self-reported motivation and steps.

\changed{It is conventional to consider a person is either an ``upward'' or ``downward'' comparer, where existing psychological research assumes that people fare better with one or the other or have stable preferences. Our MAB-based user model follows this practice and performs adaptation based on this assumption. In the meantime, data we collect from this study can help to determine whether there was enough consistency in response to one versus the other to categorize each person and then test that assumption.}

\begin{table}[t!]
	\caption{MAB configuration descriptions.}
	\begin{center}
		\begin{tabular}{|c|c|c|c|c|}
			\hline
			&\multicolumn{4}{|c|}{\textbf{Profile Steps in relation to A Participant's Steps}} \\
			\cline{2-5} 
			\textbf{Arm-}&{\bf {\em Profile 1} }&{\bf {\em Profile 2}}&{\bf {\em Profile 3}}&{\bf {\em Profile 4}} \\
			\hline
			Downward & $-40\%$ & $-30\%$ & $-20\%$ & $-10\%$ \\
			Mixed & $-20\%$ & $-10\%$ & $+10\%$ & $+20\%$ \\
			Upward & $+10\%$ & $+20\%$ & $+30\%$ & $+40\%$ \\
			\hline
		\end{tabular}
		\label{tab:mab-arms}
	\end{center}
\end{table}

% Each day and for each user, the AI selects one of these three configurations to present to the user and then observes a ``reward'' as a combination of the player's self-reported motivation at the end of the session and the player's steps observed the following day. After an introductory period (9 days) of random but equal selection among the three options, the AI engages an $\epsilon$-decreasing strategy~\cite{gray2020player} to predict which configuration will result in the highest reward for each user. The system then crafts its comparison profiles for the user based on this prediction. 

% This continues for the remainder of the study (21 days), over the course of which the AI attempts to maximize its total rewards by identifying the configuration that observes the greatest positive effect for that player. In doing so, the MAB strategy-based AI both implements an implicit player model and provides our system with the desired adaptation mechanic that caters the experience to every user.

\section{Methods} \label{sect:methods}
%\todo[inline]{Diane}
%\subsection{Participants}

To evaluate our approach of automatic personalization of social comparison, we designed a user study to test the following hypotheses: 
\begin{itemize}
    \item H1: Our MAB-based personalization mechanism is able to detect users' social comparison preference. 
    \item H2: Participants exposed to personalized social comparison will take more steps per day than those exposed to randomized social comparison. 
    \item H3: Participants exposed to personalized social comparison will report greater increase in PA motivation than those exposed to randomized social comparison. 
\end{itemize}

We recruited from a major university in the Mid-Atlantic area of the U.S. Healthy adults who had access to a {\em Fitbit} health tracker or a {\em Fitbit}-compatible smart phone were eligible if they reported PA as being somewhat or very important to them. % Individuals had to be at least 18 years of age, have daily access to a desktop or laptop computer, have access to a Fitbit account (either through a Fitbit wearable device or the Fitbit smartphone app), and provide informed consent in order to participate. 
%Participants were excluded if they had a medical condition that limited their physical activity, or if they were under medical advisement that prohibited moderate physical activity. %
Participants were compensated with extra course credits or gift cards at the end of the study. %Compensation was prorated based on participants’ level of adherence to daily activity required for participation, such that 20-21 days of adherence resulted in full compensation (2.0 SONA credits or \$15), 14-19 days resulted in partial compensation (1.5 SONA credits or \$10), and fewer than 14 days of adherence resulted in no compensation. (See Procedure section below for more information.)
%

%{\color{blue} It is worth clarifying that the primary goal of our study is to advance the understanding of how to personalize based on social comparison (H1). As a result, design decisions were made not to explicitly encourage the participants to be more active. In other words, we prioritize accurately collecting data on users social comparison tendency over directly improving their PA and motivation.} %  the game can collect more accurate social comparison behavioral data. Our concern was that if we encourage participants to compete and be more active, we will incentivize participants to display upward comparison behavior, potentially distorting their more innate tendencies. %While the results in PA are not significant, we observe positive signals. Future research can build upon what we learned about personalization and enhance it with mechanisms to encourage more PA. We have clarified our primary goal and extended our discussion to reflect this. 

\subsection{Procedure}
%Participants entered the study via rolling recruitment over the course of {\color {red} one month [Double check]}, 
At the start of the study, participants provided demographic information and completed the following baseline self-report measures. We used the Iowa-Netherlands Comparison Orientation Measure-23 (INCOM-23)~\cite{gibbons1999individual} to assess baseline social comparison tendency, which includes a 6-item upward comparison scale ({\em baseline upward score}), a 6-item downward comparison scale ({\em baseline downward score}), and an 11-item composite scale assessing general tendency towards comparison ({\em baseline composite score}). All scales have shown strong psychometric properties in previous work~\cite{gibbons1999individual}. Individuals are asked to rate how much they agree with statements describing their tendencies to make comparisons to others in various social contexts on a Likert scale (1=I strongly disagree, 5=I strongly agree). Items on the INCOM-23 include: ``I am not the type of person who compares often with others,'' and, ``I often compare how I am doing socially (e.g., social skills, popularity) with other people.'' This data was used purely for post-analysis and was not used by the AI. Since social comparison is sensitive to social contexts, we followed the convention of related studies and masked the nature of our study by collecting unrelated data (e.g., personality test).

After the baseline questionnaire, participants were asked to use our app and complete a daily survey once per day for 21 days over the course of 28 consecutive days. The daily survey asked the participants to rate their ``current motivation to exercise'' before and after their exposure to social comparison through selecting a full user profile. Responses were rated on a 5-point Likert scale (1=very low, 5=very high). 
%
%Participation occurred over the course of 28 consecutive days, in which 21 non-consecutive days of data were collected. (Four weeks of data collection were conducted to allow for occasional missing data without sacrificing opportunities for continued adherence to the three-week protocol.) 
Participants were randomized to either a control group or an experimental group using a clustered assignment to control for gender~\cite{suresh2011overview}. In the rest of the paper, the days are numbered based on the days when data is recorded, not calendar days. %All study staff involved in recruiting and coordinating participants were blind to condition assignments, as were the participants themselves.

Between Day 1 to Day 9 (i.e., the baseline period), the AI randomly selects one of the three arms with a uniform distribution (three times each), providing all participants in both groups the same exposure to all social comparison targets. 
Between Day 10 and Day 21, the control group continued viewing randomly varying profiles. The experimental group was exposed to profiles with step counts that were personalized using the AI mechanism described above. %had been selected from an adaptive user profile generator to elicit either downward or upward social comparisons. This generator attempted to automatically manipulate and optimize the social comparison environment, in order to improve motivation for exercise and increase daily step counts for those in the experimental condition.
At the end of the study, the participants completed an exit survey that collected data regarding what they found particularly valuable, whether they felt any information was missing, and their general impression. 

\changed{Of note, while other psychometrically validated measures exist to assess motivation (e.g., the Treatment Self-Regulation Questionnaire \cite{levesque2007validating}), such measures are designed to assess stable traits rather than time-dependent or momentary states within an individual. Moreover, such measures rely on retrospective recall of an individual's general tendencies, which is highly prone to bias and error. In order to assess within-person changes over time with high ecological validity (consistent with the behavioral science literature \cite{shiffman2008ecological}), the above-mentioned method to collect in-the-moment motivation data was selected.}

\subsection{Data Analysis}
%\todo[inline]{Dani}

%We evaluated the AI intervention based on four outcomes: 1)~number of observations in each MAB assignment pre-versus post-intervention (as evidence that the AI successfully differentiated players); 2)~participants' initial social comparison target selection (as evidence that the AI's assignment of participants was correct); 3)~daily change in motivation for physical activity from pre- to post-selection; and 4)~steps per day (as evidence that the AI affected real-world physical activity behavior).

%Social comparison selections were coded from 1 to 8, respectively covering the values between -40\% and +40\% depicted in Table~\ref{tab:mab-arms}. 

We used inferential statistics to analyze our quantitative data. Initial statistic analyses were performed to analyze steps and motivation changes.  %In addition, overall averages of motivation change were also analyzed directly using $t$-tests.
Because preliminary analyses showed that gender, racial identification, and age were all associated with steps per day, these characteristics were included as covariates. Average steps per day prior to introducing the personalized intervention was also included as a covariate in all analyses of step counts. Effect sizes are expressed as percent stability versus variability (target selection), semi-partial correlation coefficients (motivation for PA), and estimated step count differences between conditions.
%%Change in motivation for physical activity was calculated by subtracting post-selection motivation from pre-selection motivation. 
Differences between MAB assignments and both changes in motivation and daily steps were evaluated using two-level multilevel models with restricted maximum likelihood estimation (SAS~9.4), with days (level~1) nested within individuals (level~2). Days with <100 steps recorded were considered Fitbit non-wear days and were excluded from step count analyses (n=21). %Participants who did not complete at least 14 daily intervention sessions %were considered not to have generated sufficient data for the AI to utilize; thus, those who did not reach the predesignated minimum of 14 sessions 
%were excluded from data analyses due to insufficient data points (n=5).
Stability versus variability in selection (within-person) was evaluated using intraclass correlation coefficients (ICCs) and comparison of ICCs between conditions.

Qualitative feedback in the exit survey was coded by two independent researchers, who used grounded theory to identify themes and independently code the full dataset. They then discussed disagreements and modified their codes until they reached the inter-rater reliability of 100\%.

\section{Results}\label{sec:evaluation}
%\todo[inline]{Dani \& Gray}
%{\color{red}[Jen, could you please check that the whole Results section is written in past tense?]}
%Done.

The study enrolled 53 participants (23 male and 30 female) who were primarily Caucasian (52.8\%; Asian: 24.5\%; Multiracial: 9.4\%; African American: 7.5\%; American Indian/Native Alaskan: 1.9\%; other: 3.8\%), with a mean age of 22.45 years (\textit{SD}=7.40). Data from five participants were removed due to the lack of data points (< 14 days of completing the daily sessions). Of the remaining 48 participants, 25 (11 male and 14 female; mean age=21.80 years (SD=7.33)) were in the control condition and 23 (10 male and 13 female; mean age=23.74 (SD=8.18)) in the experimental condition.

%\subsection{AI Performance in Differentiating Participants}
%{\color{red} In Days 1-9, the AI did not employ any strategy for selecting configurations and instead offered each of the three configurations three times (in random order) to each participant. In this phase, we did not expect either the control or experimental to perform better; instead, we looked to the days of the intervention and expected that, over time, the AI would converge toward selecting either Arm-Downward or Arm-Upward for any given participant. This was observed in the $n$ columns (configuration selection counts) noted in Table~\ref{tab:step-avg-for-sessions}, where Arm-Mixed constituted only 5.7\% of all sessions in the intervention period and was never selected for any participant after Day 15. Note that throughout the intervention period, configuration selection for the control group continued to be random.} 

\begin{figure}[t!]
\includegraphics[width=0.5\columnwidth]{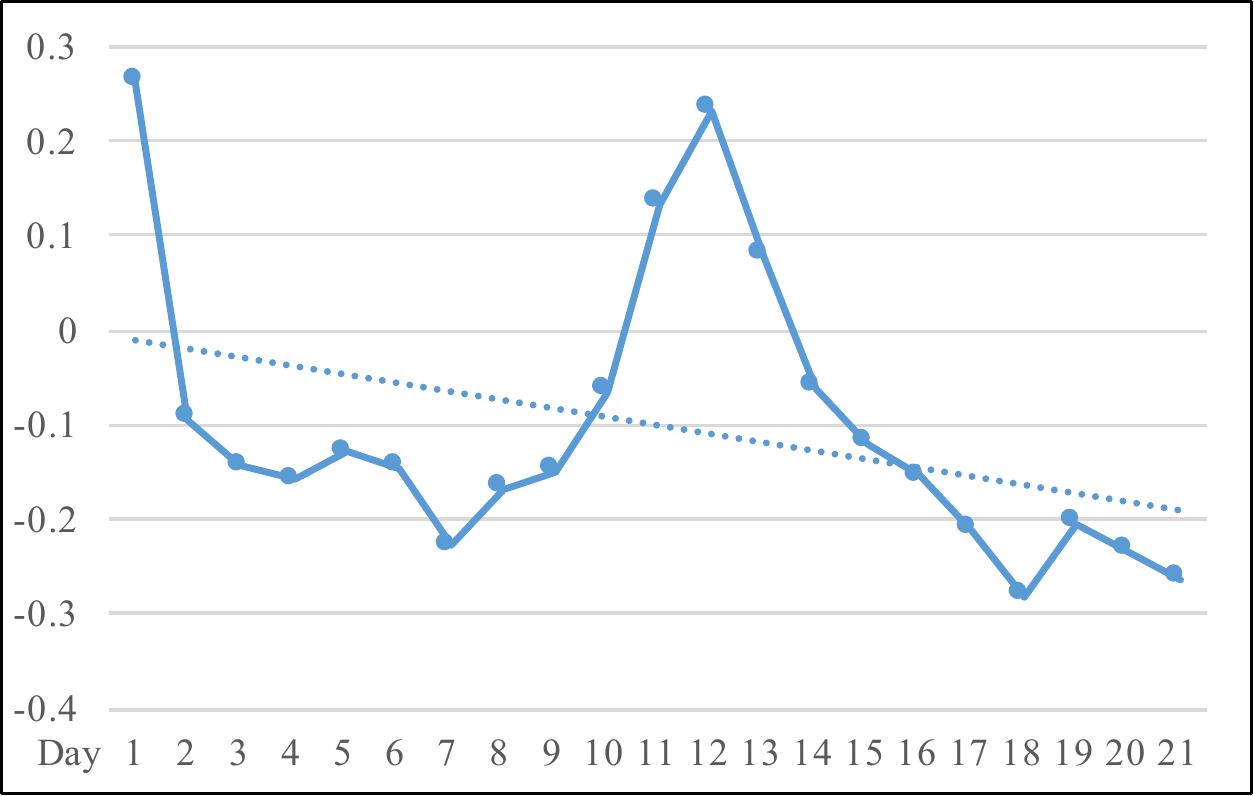}
\centering
\caption{Pearson correlation coefficient (vertical axis) between the the arm selected by the AI for each participant and their INCOM-23 scores (baseline upward score minus baseline downward score), per day. }
\label{fig:mab-incom-correlation}
\end{figure}
	
\subsection{H1: The MAB-based personalization mechanism is able to detect users' social comparison preference.}\label{subsec:results-h1}

%{\color{red} need something to compare MAB's prediction (on Day 9, Day 15, and Day 21) with INCOM-23. Need to consider the general tendency in INCOM-23.} 

We measured the correlation between our AI's predictions of individual users' comparison preference (indicated by which MAB arm was selected by the AI) and the users' self-report comparison preference (measured by their baseline upward score minus baseline downward score in the INCOM-23 instrument~\cite{gibbons1999individual}). Figure \ref{fig:mab-incom-correlation} shows the Pearson correlation coefficient (vertical axis) between the two measures throughout the 21 days (horizontal axis) for the experimental group. 
%To perform this analysis, out of all users in the experimental condition ($n=23$), we removed the 15\% of the users with lowest {\em baseline INCOM composite score} (the ones least affected by social comparison), leaving us with $n=18$ (however, we observed similar trends if all 23 users were used). 
When calculating the correlation coefficients, pairs of data points for which we had no data (e.g., if the user did not log in that day), were ignored. Given that the AI chooses MAB arms randomly in the first 9 days, we expected the correlation to be very low (close to 0) during this case. This was consistent with Figure \ref{fig:mab-incom-correlation}. %(all correlation values are between 0.1 and -0.2 in those days). 
After day 9, we expected the correlation to increase towards positive 1, as the AI learned each user's comparison preference. What we observe, however, is the opposite --- the correlation grows negative. This means that the AI learned to give users the opposite of their self-report preference measured by INCOM-23. 

To further investigate this result, we looked into the correlation between which comparison targets participants chose and their self-report comparison preference. For this purpose, we used the control group, who were not exposed to the AI intervention, and examined their comparison target choices when they were randomly assigned the Mixed arm (this is the only arm where participants were given the options to compare upward and downward) (Table \ref{tab:mab-arms}). We calculated the Pearson correlation coefficient between the control groups' self-report comparison preference (baseline upward score minus baseline downward score) and their daily target selection (-20\%, -10\%, 10\%, or 20\%). 
%We again found that our data had negative Pearson correlation (-0.2664) with the INCOM-23 scores, meaning that the users in our study tend to choose profiles {\bf opposite} to what their INCOM-23 scores seem to indicate they should. Together with the results presented in the next sections, this suggests that despite going against the INCOM-23 scores, our AI approach might be learning to show the right type of social comparison situations to the users in order to maximize their motivation to perform physical activity.
Our result has shown a close-to-zero Pearson correlation (-0.0554), indicating that INCOM-23 scores do not correlate to this group of participants' target selection.

We also calcuated the stability of participants' target selection. In both conditions, the intraclass correlation coefficient (ICC) for target selection before introducing the AI intervention was 0.00, indicating no person-level consistency in selection (100\% within-person variability between days, plus error). After the introduction of the AI intervention, however, the ICC rose to 36\% in the experimental condition versus 9\% in the control condition. Thus, stability in target selection was greater in the experimental condition after the introduction of the AI intervention ($\chi^2$ = 8.20, p = 0.02). This means that the AI constrained participants’ selection options to a narrower range than random assignment.

\subsection{H2: Participants exposed to personalized social comparison will take more steps per day than those exposed to randomized social comparison. }

Average step counts per day by condition and MAB configuration appear in Table~\ref{tab:step-avg-for-sessions}. A modest average decrease in steps was observed across both conditions when aggregating pre- and post-intervention steps. This trend was consistent with previous research on mobile health interventions that do not contain explicit PA goals, in which novelty effects and increased awareness of one's PA may influence behavior at baseline, prior to habituation \cite{shin2019beyond}.

\begin{figure}[t!]
\includegraphics[width=0.75\columnwidth]{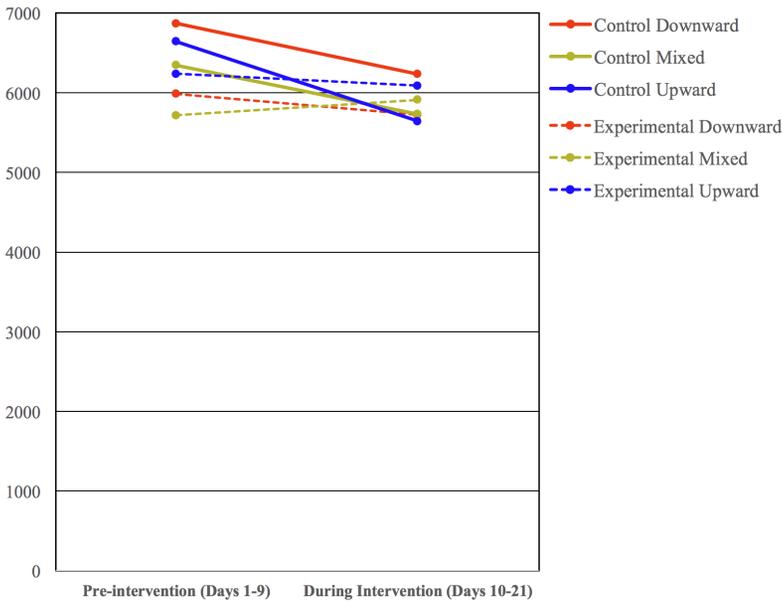}
\centering
\caption{Visualization on the average step data reported in Table \ref{tab:step-avg-for-sessions}.}
\label{fig:step-avg-plot}
\end{figure}

We can see some interesting patterns after the introduction of the AI. For example, when comparing the average step count of participants that were given Arm-Downward between days 1-9 and days 10-21, we see that participants in the control group went from 6869 steps on average to 6234 (a decline of 635). In contrast, the participants in the experimental condition changed from 5987 to 5722 (a decline of only 265). The same trend can be observed for the other arms. For Arm-Mixed, the control group shows a decline of 615 steps, while the experimental group shows an {\em increase} of 195 steps. For Arm-Up, the control group decreases in 1000 steps, while the experimental group only dropped 150. Fig. \ref{fig:step-avg-plot} visualizes the changes between average steps in Table \ref{tab:step-avg-for-sessions}. The solid lines represent the trend in average steps in the control group observed for Downward/Mixed/Upward respectively, while the dashed lines correspond to those seen in the experimental group. As the plot shows, the introduction of the AI visibly altered the downward trend seen in the control group. 
%{\color{red}Is this statistically significant?}

However, between the two conditions, we did not observe any statistical difference in their steps during intervention period. While the interaction of condition and MAB assignment was non-significant, we found an interesting pattern of daily steps between conditions and MAB arm assignment. %but estimated step counts between condition, MAB, and study period had interesting patterns. 
During intervention period, those who received the MAB upward arm in the experimental condition took an appreciably larger number of steps per day than those who were randomly assigned the same MAB upward arm in control. The pattern reversed with downward arm assignment, where control group participants who received this assignment took more steps per day than those in the experimental group. 

%\red{add results of testing for cumulative effect. Either here or in Discussion}\todo[inline]{Dani: please check your data analysis.}

\begin{table*}[t!]
	\centering
	\caption{Step averages for sessions in control and experimental groups.}
	\label{tab:step-avg-for-sessions}
	\resizebox{\textwidth}{!}{% <-- do not delete this %
	\begin{tabular}{|l|l|l|l|l|l|l|l|l|}
		\multicolumn{1}{c}{} & \multicolumn{4}{c}{\textbf{Pre-Intervention (Days 1-9)}} & \multicolumn{4}{c}{\textbf{During Intervention (Days 10-21)}} \\ \cline{2-9} 
		\multicolumn{1}{l|}{\textbf{Arm}} & \multicolumn{2}{c|}{\textit{\textbf{Control}}} & \multicolumn{2}{c|}{\textit{\textbf{Experimental}}} & \multicolumn{2}{c|}{\textit{\textbf{Control}}} & \multicolumn{2}{c|}{\textit{\textbf{Experimental}}} \\ \cline{2-9} 
		\multicolumn{1}{l|}{} & \multicolumn{1}{c|}{\textit{n}} & \multicolumn{1}{c|}{\textit{S}} & \multicolumn{1}{c|}{\textit{n}} & \multicolumn{1}{c|}{\textit{S}} & \multicolumn{1}{c|}{\textit{n}} & \multicolumn{1}{c|}{\textit{S}} & \multicolumn{1}{c|}{\textit{n}} & \multicolumn{1}{c|}{\textit{S}} \\ \hline
		\multicolumn{1}{|l|}{Downward} & \multicolumn{1}{l|}{73} & \multicolumn{1}{l|}{6869 (SE=767)} & \multicolumn{1}{l|}{69} & \multicolumn{1}{l|}{5987 (SE=714)} & \multicolumn{1}{l|}{91} & \multicolumn{1}{l|}{6234 (SE=549)} & \multicolumn{1}{l|}{100} & \multicolumn{1}{l|}{5722 (SE=552)} \\
		\multicolumn{1}{|l|}{Mixed} & \multicolumn{1}{l|}{75} & \multicolumn{1}{l|}{6345 (SE=765)} & \multicolumn{1}{l|}{69} & \multicolumn{1}{l|}{5717 (SE=714)} & \multicolumn{1}{l|}{79} & \multicolumn{1}{l|}{5730 (SE=571)} & \multicolumn{1}{l|}{13} & \multicolumn{1}{l|}{5912 (SE=588)} \\
		\multicolumn{1}{|l|}{Upward} & \multicolumn{1}{l|}{74} & \multicolumn{1}{l|}{6643 (SE=767)} & \multicolumn{1}{l|}{67} & \multicolumn{1}{l|}{6238 (SE=717)} & \multicolumn{1}{l|}{98} & \multicolumn{1}{l|}{5643 (SE=548)} & \multicolumn{1}{l|}{109} & \multicolumn{1}{l|}{6088 (SE=492)} \\ \hline
		
		Total & 222 & 6617 & 205 & 5978 & 268 & 5869 & 222 & 5913 \\ \hline
	\end{tabular}% % <-- do not delete this %
	}
\end{table*}

\subsection{H3: Participants exposed to personalized social comparison will report greater increase in PA motivation than those exposed to randomized social comparison. } 

Table \ref{tab:least-squares-means} summarizes the differences in participants' motivation to exercise between pre- and post-selection. The bottom two lines of the table show the average $\Delta$ motivation for the control and experimental group. While the change in motivation in the control group was just 0.0194, the experimental group reported a larger change in motivation of 0.1456. This difference was found to be statistically significant using a t-test ($p = 0.0038$).

However, when we considered the demographic differences, the difference between conditions is less significant. Controlling for gender, racial identification, and age, motivation for PA across the AI-based personalization intervention period increased from pre- to post-selection in the experimental condition (B = 0.08, SE = 0.11). In contrast, motivation decreased symmetrically in the control condition (B = -0.08, SE = 0.12). Across condition and MAB assignments, change in motivation was most positive among participants in the experimental condition when assigned Arm-Mixed (mix of targets). Although the difference in motivation change between conditions did not reach statistical significance (F = 1.85, p = 0.17), it was associated with a small-to-moderate effect size (sr = 0.20) in favor of the experimental condition.

\begin{table}[t!]
	\centering
	\caption{Difference in motivation pre- to post-selection.}
	\label{tab:least-squares-means}
	\begin{tabular}{|llll|}
		\multicolumn{4}{c}{Least Squares Means} \\ \hline
		\multicolumn{1}{|l}{Condition} & MAB Assign. & $\Delta$ Motivation & \multicolumn{1}{l|}{Std. Err.} \\ \hline
		\multicolumn{1}{|l}{Control} & Arm-Downward & 0.0071 & \multicolumn{1}{l|}{0.1268} \\
		\multicolumn{1}{|l}{Control} & Arm-Mixed & -0.1189 & \multicolumn{1}{l|}{0.1315} \\
		\multicolumn{1}{|l}{Control} & Arm-Upward & -0.1117 & \multicolumn{1}{l|}{0.1258} \\
		\multicolumn{1}{|l}{Experimental} & Arm-Downward & 0.1013 & \multicolumn{1}{l|}{0.1267} \\
		\multicolumn{1}{|l}{Experimental} & Arm-Mixed & 0.3099 & \multicolumn{1}{l|}{0.2216} \\
		\multicolumn{1}{|l}{Experimental} & Arm-Upward & 0.0395 & \multicolumn{1}{l|}{0.1134} \\ \hline 
		\multicolumn{4}{c}{Overall Means} \\
		\hline
		Control & - & 0.0194 & 0.0201 \\
		Experimental & - & 0.1456 & 0.0400 \\ \hline
	\end{tabular}
\end{table}

\subsection{Qualitative Results}
\label{QR}
%\todo[inline]{Jen, Diane}

Post-study survey responses were provided by 40 out of 53 participants. Qualitative feedback was coded into 22 distinct themes that arose related to profile features that participants found to be either ``particularly valuable or interesting'' or ``missing'' from the website. Overall, the tone of the qualitative feedback suggested that participants believed the profiles belonged to real people and that they valued the information provided.%, which supported our assertion that genuine social comparisons were performed.

When asked to describe features of the profiles that they found most valuable or interesting, participants identified the profiles' step counts (35\%), occupations (35\%), and hobbies (25\%) to be most interesting to examine. For example, one participant wrote that they were most interested in the profiles' ``{\em occupation and hobbies/interests. It was interesting to put the number of steps into perspective.''} 

A minority of participants (35\%) explicitly acknowledged that they performed social comparison. They had compared themselves to the artificial user profiles regarding users' interests (10\%), exercise routines (10\%), and steps (10\%). One participant wrote: ``{\em I found that looking at... the interests was valuable as well, since it helps you compare yourself to the other people and see how maybe someone similar to you works out.}'' Still, others (5\%) reported having made explicit comparisons to the artificial users' demographic data and physical characteristics. ``{\em It was most helpful to look at people in a similar demographic as me (sex, age, and weight wise).}''

When asked to describe features of the profiles that were missing, many (20\%) reported that they would have liked to see more detailed information regarding specific exercise routines. For example, one participant wrote: ``{\em When it comes to exercise, it would have been interesting to see how different people work out and what they specifically do for fitness.}'' Others noted they would have liked to know more about physical characteristics (e.g., weight history, 7.5\%) and psychological well being (e.g., mood, stress, 12.5\%) in the user profile.

\section{Discussions}
This section provides our interpretation of the results and discusses the design implications of social comparison-based features in fitness apps. 

\subsection{Automatic Model of Individuals' Social Comparison Preference}

\begin{itemize}
    \item H1: The MAB-based personalization mechanism is able to detect users' social comparison preference.
\end{itemize}

Overall, we observed some evidence that our AI-based personalization mechanism was able to model users' social comparison preferences. First, following existing literature, we assume that a user's social comparison preference is either upward or downward. This domain knowledge is not encoded in the AI. As shown in Table~\ref{tab:step-avg-for-sessions}, there is a significant drop in the number of times the AI chose the Arm-Mixed option in the experimental group, an option that indicates the AI's uncertainty. This indicates that the AI was able to converge on either Arm-Downward or Arm-Upward and differentiate participants based on their daily data. The stability in participants' comparison target selection, measured by the intraclass correlation coefficients (ICCs), increased more in the experimental condition than in the control condition. In addition, participants' target selections increasingly aligned with their MAB assignment. However, this is not conclusive evidence, since this phenomenon may be partially caused by the narrowing effect of the AI algorithm to provide social comparison targets of the same direction. %Our measurement does not control for the fact that the opportunities for selection were purposely limited in the experimental condition by the function of Arm-Downward and Arm-Upward; however, we argue that consistency in participant selection could serve as a proxy for player satisfaction. %In other words, we hypothesize that a player dissatisfied with their placement in the model would have more erratic choices. This may be investigated in future studies.

A main source of challenge to test H1 is related to measuring participants' self-report social comparison preference using the INCOM-23 instrument. We did not observe correlation between participants' INCOM-23 scores, their actual target selections, and the AI algorithm's MAB arm assignment (which is the user model of one's social comparison preference). 

%{\color{red}  Together with the results presented in the next sections, this suggests that despite going against the INCOM-23 scores, our AI approach might be learning to show the right type of social comparison situations to the users in order to maximize their motivation to perform physical activity.}

%Overall, we did not observe any {\color{blue} positive} correlation between users' self-report social comparison preferences and their behavior, {\color{blue} and, in fact, we observed a slight negative correlation (Section \ref{subsec:results-h1}}. 

% MAB strategies are typically evaluated ``in the limit'' and functionally expect a number of observations orders of magnitude higher than the 21 provided in this study. We consider the fact that the AI was able to converge on either Arm-Downward or Arm-Upward for all but 5.7\% of sessions after only 9 observations per player as a successful demonstration of the low-horizon bandit techniques we employed \cite{gray2020player_blind}.

%the decreasing counts for Arm-Mixed reflected in Table~\ref{tab:step-avg-for-sessions} indicate that the MAB was successful in differentiating participants.
%
%Furthermore, users' selections of their social comparison targets are more stable among the experimental group than the control group. It may suggest that users in the experimental group were provided with the social comparison targets they prefer. {\color{red}This could also suggest that the users don't really have a choice. Think more.} 

This could be the result of several factors. First, participants may be selecting comparison targets not only based on their steps but also on other factors. Although we purposefully designed the first two selection pages of our app to focus primarily on step information and minimize demographic information (e.g., using non-descriptive user names such as ``azb30''), it is possible some participants chose comparison targets also based on the other available information in the Overview Page (exercise location, favorite spots, average distance, or interest). \changed{However, our qualitative results indicate that this influence is limited. To the question of which user profile information was the most valuable or interesting, 60\% of the participants reported occupations or hobbies compared to 35\% who said step counts. However, users can only access the occupation or hobby data {\em after} making their final selection of user profiles. Furthermore, although our log files show that 81.13\% of the participants %(88.89\% in the control group and 72.08\% in the experimental group) 
selected more than one profile to preview in at least one daily session, only 12.07\% of the total sessions involved participants previewing more than one profile. For both conditions, most of these sessions (68.00\% for the control group and 76.74\% for experimental) happened in the first 9 days, when the app was novel. This suggests that while different demographic information was meaningful to the participants, it did not substantially impact how the participants selected their comparison targets. In the majority of the sessions, the participants chose their comparison targets outright, primarily based on step count information.}

Second, \changed{an individual's social comparison tendency may have more variability than existing literature suggests. Only recently has the potential variability of social comparison preference and response received considerable attention~\cite{arigo2019methods}. Given that the preponderance of literature treats social comparison as a stable individual difference, our MAB-based AI model makes the same assumption so that we can make progress on personalization technology and design. In addition, our choice of an MAB-based approach enabled our AI to robustly handle some variability in users’ behavior and pick up the overarching trend. It is possible that our web app was able to pick up the daily variances over the course of the intervention, while the INCOM-23 survey only measured participants' overall tendencies once at the beginning of the study. It is likely that the INCOM picks up participants' {\em perceptions} of their preferences and behavior in aggregate. Furthermore, this self-report is likely infused with social desirability biases and misrecollection to a degree that actual social comparison {\em behavior} is not. 
In the meantime, we designed our personalization system for repeated exposure (i.e., 9 times/days), rather than assessing preference/response on one occasion as in previous work~\cite{mollee2016effectiveness}. This is a step toward addressing this variability. Our findings indicate that some participants had varying responses and thus contribute new evidence to psychological research on social comparison, the foundation of many social comparison-based features in social applications.}

\subsection{Personalized Social Comparison and PA Changes}

\begin{itemize}
    \item H2: Participants exposed to personalized social comparison will take more steps per day than those exposed to randomized social comparison. 
\end{itemize}

We observed an overall decrease in the number of steps from days 1-9 to days 10-21, which is not unexpected given the nature of this sample. Individuals who had a casual interest in physical activity were not necessarily strongly motivated to increase their activity levels.  In terms of the effect of condition, we observed that participants in the control group had a steep decline in steps from 1-9 to 10-21, while participants in the experimental condition experienced a smaller decline (Figure \ref{fig:step-avg-plot}); however, the effect of condition did not reach statistical significance. 
%participants’ steps per day did not differ overall between conditions. However, while statistically nonsignificant, meaningful differences in daily steps were observed in the interaction between condition and MAB assignment, such that those in the experimental group assigned to Arm-Upward took several hundred more steps per day than those with the same assignment in the control group. Meanwhile, this pattern was reversed for those assigned to Arm-Downward, such that those in the control group took more steps than those in the experimental group. Thus, there appeared to be meaningful differences between Arm-Upward and Arm-Downward, where Arm-Upward worked as intended while Arm-Downward did not. These results may be explained by the notion that social comparison does not always result in improved performance; rather, the impetus for comparison may be self-evaluation or self-enhancement~\cite{Wood1989}, rather than self-improvement. 

If a need for self-enhancement drove a participant's comparisons, that participant may have chosen targets of comparison that would aid them in ``protecting or enhancing [their] self-esteem,'' for which downward comparisons are likely more effective ~\cite{Wood1989}. We might, therefore, expect effective downward comparisons to result not in higher performance, but rather increased positive affect~\cite{arigo2019methods}. While the present study did not measure participant affect following completion of daily sessions, we did gather a self-reported measure of motivation, which exhibited an increase in the Arm-Downward participants that is more than three times greater than that in the Arm-Upward participants. Future research in this area should consider evaluating momentary affective or emotional responses following social comparison in order to empirically evaluate this question.

\begin{itemize}
    \item H3: Participants exposed to personalized social comparison will report greater increase in PA motivation than those exposed to randomized social comparison. 
\end{itemize}

With respect to the daily change in motivation, participants’ motivation increased among those in the experimental group and decreased within the control group. While we observed some  indications of difference between conditions, after controlling for gender, race, and age, the difference between control and experimental conditions was not statistically significant. However, these differential outcomes by condition, along with the small-to-moderate effect sizes observed, indicate that there was a positive effect of the AI personalization on motivation. 

Moreover, the biggest effect was seen in the
%However, this effect appears to have been driven by 
assignment to Arm-Mixed within the experimental group, for which there were far fewer observations, and in which randomly selected targets were displayed rather than targets tailored to match participants’ social comparison preferences. Improved motivation in Arm-Mixed may have been observed because participants were able to explore a wider array of targets beyond those solely matching their preferences, which could have ignited greater interest in differentiating between targets to find the option most aligned with their preference. This is consistent with previous work showing that individuals exposed to a range of comparison targets explore several different types of targets but spend considerable time reviewing information about a smaller subset~\cite{van2000social}.

\changed{Qualitative findings indicate that participants largely found the information contained in the user profiles to be valuable. While participants did not directly report on the impact of their interest in the profiles on their motivation for PA, several participants described features that led them to make social comparisons. Among the participants who explicitly acknowledged that they performed social comparison, they were evenly split between upward and downward comparers according to our MAB user model. This suggests that both types of participants were engaged with their comparison targets provided by the app, which is the requirement for activating the psychological process of social comparison.}

%\red{[may need to move this elsewhere. The Methods section?] Another experimental designs is to introduce a condition where the app provides comparison targets the opposite of user preference. However, evidence in psychology indicates this is demotivating [add citations]. As the goal of this study is to model individual comparison preference, we believe a random baseline is a more suitable benchmark.} \todo[inline]{Dani: please add relevant citations here.}

In summary, the results of the present study indicate that our AI-based personalization approach was able to automatically model and manipulate social comparison in the pursuit of PA promotion. Although methodological limitations (described below) preclude a definitive conclusion that the intervention successfully motivated PA, the detected effects achieved small-to-moderate effect sizes, illustrating the real-world implications of the intervention for enhancing motivation and daily steps. Moreover, results from our qualitative analyses suggest that participants valued and applied the information provided in the profiles to make social comparisons. These findings suggest that future work, which can incorporate the present methodology into engaging modalities such as online games and social networks, may be able to harness the effects of AI-based personalization to enhance motivation and PA.

\subsection{Design Implications}
Since social comparison is frequently used in m-health apps for PA and \changed{in social media applications}, we summarize the following key design lessons we learned in this project. 
%What are our interpretations about whether our system biased users' ``natural'' social comparison tendency? Anything else? 

First, the relatability of social comparison targets is critical. In our UI/UX design, we developed a large number of artificial user profiles and made efforts to ensure all content was realistic and varied across profiles. Our qualitative analysis showed how participants engaged with user profiles they can relate to. Participants found it ``more helpful'' to compare themselves with people who were ``in similar demographic'' and see how ``... someone similar to you works out.'' This suggests that when designing social comparison features, independent from comparison direction, it is beneficial to provide users with comparison targets they can relate to. In addition to targets with similar demographics, another known phenomenon is that users of social m-health apps relate more closely to people with whom they have a positive prior relationship (e.g., friendship and peer relation)~\cite{salvy2009effect,fitzgerald2012peers,caro2018understanding}.  

Second, it is important to balance the system's need for data collection and usability. Building an accurate real-time user model of social comparison requires collecting relevant data. As users interact with a web app, it can be difficult to track which information on the page they are paying attention to. To solve this, we purposefully designed our UI to require more steps in user interaction. For instance, modern web UI/UX design may streamline all user profile information we provide into a single page; however, doing so will significantly reduce the app's ability to capture important user behavior that can be used to indicate the user's attention. Given the special data-gathering requirement of AI-based personalization systems, we recommend designers re-evaluate the balance between usability requirements (e.g., efficiency) and data-gathering based on the specifics of their projects. 

\changed{Third, perhaps the most important, further research is needed on the {\em personalization paradox}~\cite{zhu2020Personalization}. We will discuss this in the following section. }

\changedcamera{
\subsection{Personalization Paradox}}
Jarno Koponen was the first to coin the phrase ``{\em personalization paradox}.'' In a 2015 {\em TechCrunch} article,\footnote{Retrieved from \url{https://techcrunch.com/2015/06/25/the-future-of-algorithmic-personalization/}, Oct 1, 2020.} he wrote 
\begin{quote}
``{\em [There] lies a more general paradox at the very heart of personalization. Personalization promises to modify your digital experience based on your personal interests and preferences. Simultaneously, personalization is used to shape you, to influence you and guide your everyday choices and actions. Inaccessible and incomprehensible algorithms make autonomous decisions on your behalf. They reduce the amount of visible choices, thus restricting your personal agency.}''
\end{quote}
While it is an insightful observation, the discussion around the personalization paradox has so far remained abstract. Reflecting on our work above, we further argue that the personalization paradox is a result of the fundamental {\bf conflict between user modeling and adaptation}. Below are the two main ways they can be at odds with each other (notice these are related and are not mutually exclusive): 

\begin{enumerate}
\item \changedcamera{ {\bf The Self-Reinforcing Loops Problem} happens when a personalization system ``forces'' a user into what its user model categorizes, regardless of whether the model is accurate. This self-reinforcing nature of personalization technology has been documented by other researchers~\cite{o2016weapons,noble2018algorithms,pariser2011filter}. For instance, imagine the {\em Netflix}'s user model inaccurately predicts a specific user's preference to be only Sci-fi based on her viewing history and thus only recommends Sci-fi content. Since the user can only express her preference through the digital environment controlled by the personalization algorithm, she is more likely to further display Sci-fi preference due to the lack of other choices. In this way, the adaptation reinforces its user model without a chance to adjust the latter.}

\item \changedcamera{ {\bf The Moving Target Problem} happens when the user changes with the digital environment she is in. %Even when no self-reinforcing loops occur, the fact that personalization affects the user's behavior is still problematic. 
Continuing the above example, let us assume that the user model correctly categorized a user's preference to Sci-fi when she is in the ``neutral'' viewing environment with a wide range of different genres. However, when adaptation modifies this context into a Sci-fi-heavy one, the user's preference may change to, say, documentary. This case is the closest to Kopenen's description above --- by changing the users' digital environment through adaptation, users' preferences and behavior become a moving target for modeling. A special case of this problem occurs when the goal of the personalization system is to induce behavior change. In this situation, the system's explicit goal is to push the user's preferences or behavior in a particular direction. As a result, user modeling might reflect the user at the start rather than what she has become.}
\end{enumerate}

\changedcamera{Reflecting on our project, we attempted to minimize the {\em self-reinforcing loops problem} by including the {\em Mixed} arm in our MAB algorithm and thus allowed users to choose upward or downward profiles. However, this does not completely address the problem. Self-reinforcing loops could occur if the MAB is inaccurate but very confident. In these cases, the user would be given either all {\em Upward} or {\em Downward} arms, leaving no opportunity for her to express a different preference. A potential solution to address this problem is for the MAB algorithm to force explore (e.g., always use the {\em Mixed} arm) every once a while. 
%we believe that we successfully avoided the {\em self-reinforcing loops problem}. By always giving users at least one comparison target different from their user model, we have provided an option for them to ``correct'' the user model at any time. %We also designed the platform to be neutral in order to minimize biasing users towards any comparison directions (upward or downward). 
}

\changedcamera{Our attempt to minimize the {\em moving target problem} was also met with mixed success. It is unclear whether being more physically active impacts one's comparison tendency (e.g., upward). For instance, when a user increases her PA, will she compare upward more often? Currently, psychology literature does not have a definitive answer. If so, the moving target problem of the changing user in our case may be exacerbated by the targeted behavior change in addition to the adapted digital environment. In order to mitigate this risk, we designed the system to be as neutral as possible; for instance, we intentionally left out the design elements (e.g., competition) which are known to increase engagement and PA but risk pushing users toward upward comparison and penalizing downward comparison. However, this may have significantly reduced the app's ability to motivate changes in PA.  Conversely, if we had decided to incorporate features known to motivate PA effectively, we risk skewing users' preferences and thus jeopardizing the accuracy of the user model.
}

\changed{To mitigate the personalization paradox, we believe that a project's first step is for the AI engineers/data scientists and designers/behavioral scientists to have an open conversation about the priorities of the project. In our case, for example, given the lack of work in this area, prioritizing model accuracy was appropriate. Another approach is to further separate the user modeling and adaptation stages of personalization. For example, our system initially remained neutral to collect accurate user data on social comparison tendencies. Once it had a robust model, it could then unlock other design features to explicitly motivate PA. However, more research is needed to see if users' social comparison tendencies in a neutral environment can be transferred into a different context with new features that may reward upward comparison.  As this research matures, we may want to strike a different balance to have stronger behavioral outcomes. As personalization becomes increasingly popular in areas associated with behavior change, such as health and education, further research is needed to balance personalization model accuracy and behavioral change effectiveness.
}

\subsection{Limitations}
%\todo[inline]{Diane, Gray}

%The present study had significant strengths. This study represents the first application of social comparison theory that avoided common assumptions about social comparison behaviors. First, the study design acknowledged that a one-time assessment of comparison behavior is not sufficient or generalizable, as behaviors and preferences vary frequently within each person; rather, repeated measurements of motivation and physical activity were taken to ensure accurate representation of social comparison behaviors and preferences within each individual. Second, the present study used objective assessment of daily steps, rather than the self-report measures commonly used that allow for possible retrospective recall errors. Finally, we employed a multilevel statistical approach to maximize the AI's learning from repeated measures. These innovative technological and analytic tools enabled us to capture, model, and evaluate participants' social comparison behaviors and preferences, daily activity, and daily changes in motivation with greater accuracy than previous work.

The present study had a few limitations. First, due to the relatively small sample size, our analyses were underpowered to detect the proposed effects with statistical significance. %(Of note, the study ceased recruitment and enrollment prior to achieving full power due to the constraints of recruiting a convenience sample of undergraduate students on an academic calendar.)
%Thus, effect sizes were primarily used to interpret results based on the magnitude of each effect, in order to evaluate the real-world implications of the study's results. Even still, 
Results thus must be interpreted with some caution, and further research with larger samples is needed.
Second, participants were able to complete sessions on the web application at any time of day during each day of participation. Because their access to the web app was not restricted to a specific, standardized window of time each day, the time of exposure to social comparison targets was not necessarily consistent within or between participants (e.g., mid-morning for one participant vs. late evening for another). As a result, the intervention's effects on steps may have been spread across days for some individuals, rather than having consistently had impact on the same day, as our analyses assumed. Third, the age range of our participants was relatively narrow, which limits the generalizability of our results to a broader age range. %A further limitation was the use of a sample of convenience who were not strongly motivated to increase their physical activity.  As such, results may not generalize to those who are strongly motivated to increase their physical activity; one likelihood is that the effects would be stronger amongst that group. 
%Future research should enforce standardized session times such that the temporal effects of the intervention can be measured with greater confidence.
%
%Third, the AI did not take into account the subjective target preferences of the participants (i.e., the choices made among the four profiles presented each day), and instead reinforced its mapping between the {\em opportunities} presented to the participants and the behavioral outcomes following exposure to those opportunities (e.g., daily steps).  However, consideration of participants' selections may reveal observable effects \cite{mollee2016effectiveness} that could help to maximize the effectiveness of the AI for capturing social comparison behavior. Future work should investigate how this might be integrated into the reinforcement loop of the MAB.
%
\changed{Finally, we made the assumption that users' social comparison preference is static based on established literature; however, this understanding has started to be challenged in recent work in psychology. Future research is needed to validate this assumption and evaluate whether a dynamic user modeling AI technique would be appropriate.} 
%\todo[inline]{Jichen: add we do not focus on ``PA goals.'' ``When the current design incorporates profile information, whether these information will have a possible impact on the user needs to be further clarified.'' A simplified (but consistent with the field) notion of SC trends? Steps is a rudamentary measure for PA, future work can look into differentiate moderate or moderate-to-vigorous exercise. }

\section{Conclusions and Future Work}
In conclusion, we presented our approach to personalize social comparison, the basis of many widely adopted features in both commercially available m-health apps and research prototypes for promoting physical activity. Our work is among the first to investigate how to personalize social comparison automatically for social fitness apps. Results of our user study indicate that our AI-based personalization approach was able to automatically model and manipulate social comparison in the pursuit of PA promotion. The detected effects achieved small-to-moderate effect sizes, illustrating the real-world implications of the intervention for enhancing motivation and daily steps. We also proposed design guidelines for future social comparison-based m-health applications. 

As part of our future directions, we plan to evaluate our approach with profiles from real users, especially among users with existing positive social relationships. We also plan to incorporate the present methodology into engaging modalities such as social mobile games to further harness the effects of AI-based social comparison personalization to enhance motivation and physical activity. Last but not least, we intend to investigate how to solve the personalization paradox and make our approach more effective. 

\section{Acknowledgement}
\changedcamera{This work is partially supported by the National Science Foundation (NSF) under Grant Number IIS-1816470. The authors would like to thank all past and current members of the project. }

% BALANCE COLUMNS
% \balance{}

% REFERENCES FORMAT
% References must be the same font size as other body text.

% \newcommand{\showDOI}[0]{\unskip}

\bibliographystyle{SIGCHI-Reference-Format}
\bibliography{references}

\end{document}